\documentclass[preprint,12pt]{elsarticle}
\usepackage{amssymb}
\usepackage{amsmath}

\makeatletter
\renewcommand{\ps@pprintTitle}{
	  \renewcommand{\@oddfoot}{}
	  \renewcommand{\@evenfoot}{}
	}
\makeatother

\begin{document}
	
\begin{frontmatter}
		
		\title{Machine-Learning-Assisted Pulse Design for State Preparation in a Noisy Environment} 
		
		\author{Zhao-Wei Wang$^{a}$, Hong-Yang Ma$^{b}$, Yun-An Yan$^{c,*}$, Lian-Ao Wu$^{e,f,g}$ and Zhao-Ming Wang$^{a,d,\dagger}$} 
		
		\affiliation[label1]{organization={College of Physics and Optoelectronic Engineering},
			addressline={Ocean University of China},
			city={Qingdao},
			postcode={266100},
			state={China}}
		
		\affiliation[label2]{organization={School of Sciences},
			addressline={Qingdao University of Technology},
			city={Qingdao},
			postcode={266033},
			state={China}}
		
		\affiliation[label3]{organization={School of Physics and Optoelectronic Engineering},
			addressline={Ludong University},
			city={Shandong},
			postcode={264025},
			state={China}}
		
		\affiliation[label4]{organization={Engineering Research Center of Advanced Marine Physical Instruments and Equipment of Ministry of Education},
			addressline={Ocean University of China},
			city={Qingdao},
			postcode={},
			state={China}}
		
		\affiliation[label5]{organization={Department of Physics},
			addressline={University of the Basque Country UPV/EHU},
			city={Bilbao},
			postcode={48080},
			state={Spain}}
		
		\affiliation[label6]{organization={IKERBASQUE, Basque Foundation for Science},
			addressline={},
			city={Bilbao},
			postcode={48013},
			state={Spain}}
		
		\affiliation[label7]{organization={EHU Quantum Center},
			addressline={University of the Basque Country UPV/EHU},
			city={Leioa},
			postcode={48940},
			state={Biscay, Spain}}
		
		\begin{abstract}
			High-precision quantum control is essential for quantum computing and quantum information processing. However, its practical implementation is challenged by environmental noise, which affects the stability and accuracy of quantum systems. In this paper, using machine learning techniques we propose a quantum control approach that incorporates environmental factors into the design of control schemes, improving the control fidelity in noisy environments. Specifically, we investigate arbitrary quantum state preparation in a two-level system coupled to a bosonic bath. We use both Deep Reinforcement Learning (DRL) and Supervised Learning (SL) algorithms to design specific control pulses that mitigate the noise. These two neural network (NN) based algorithm both have the advantage that the well trained NN can output the optimal pulse sequence for any environmental parameters. Comparing the performance of these two algorithms, our results show that DRL is more effective in low-noise environments due to its strong optimization capabilities, while SL provides greater stability and performs better in high-noise conditions. These findings highlight the potential of machine learning techniques to enhance the quantum control fidelity in practical applications.
		\end{abstract}
		
		
\end{frontmatter}

\section{Introduction}

As quantum information science advances rapidly, quantum control technology has become a key enabler of breakthroughs in quantum computing \cite{nielsen2010quantum, zhang2022quantum, cho2021quantum} and quantum communication \cite{cariolaro2015quantum, cozzolino2019high}. Although quantum control schemes perform well in noise-free conditions \cite{wu2009perfect, wang2020almost, zhang2019does, he2021deep, bartkowiak2014quantum}, real-world applications are inevitably affected by interactions with external environmental noise \cite{torovs2021relative,de2021materials, wang2020nonMarkovian, barenco1997effects, benedetti2013dynamics, paladino20141, tchoffo2016quantum}. Noise not only reduces the stability and control accuracy of quantum systems but also causes quantum information loss, significantly hindering the practical application of quantum technology.

Quantum state preparation (QSP) is a core task in quantum information processing, with the goal of transforming an initial quantum state into a target quantum state. This process is of great significance for the initialization of quantum systems and the execution of quantum algorithms. For example, in the implementation of quantum Fredkin or Toffoli gates, it is necessary to pre-prepare the ancilla state to the standard state $ |0\rangle $ in certain cases \cite{PhysRevA.53.2855,PhysRevA.51.1015}. When implementing the Variational Quantum Eigensolver algorithm, it is often necessary to prepare the quantum circuit in the Hartree-Fock state\cite{TILLY20221}. When transferring a quantum state through a spin chain, the initial state of the system usually needs to be set to the ground state \cite{wang2020almost}. Moreover, in completing the task of quantum teleportation, two-qubit entangled states must be generated in advance\cite{nielsen2010quantum}. However, when the preparation task changes, it is usually necessary to redesign the control strategy. Therefore, when facing the preparation tasks of multiple initial states and multiple target states, i.e., arbitrary quantum state preparation(AQSP), the design of control strategies can become extremely complex. In addition, the presence of noise and decoherence in real quantum systems further increases the difficulty of this task, as the control strategy must be robust against these imperfections.

Various strategies have been proposed to reduce the impact of noise on quantum dynamics, including the use of quantum error correction codes \cite{roffe2019quantum, knill1997theory, laflamme1996perfect, calderbank1996good}, dynamic decoupling \cite{khodjasteh2005fault, jing2014nonperturbative, mena2022protectability}, and dynamically corrected gates \cite{yang2018neural, wang2012composite, khodjasteh2010arbitrarily, khodjasteh2012automated, wan2016free}. 
Quantum error correction codes, such as the Bacon-Shor code \cite{aliferis2007subsystem} and the surface code \cite{fowler2012surface}, encode logical qubits into entangled states of multiple physical qubits. This encoding enables the detection and correction of errors without directly measuring the logical qubits, thereby preserving the integrity of quantum information.
Dynamic decoupling \cite{wang2012nonperturbative} suppresses noise by applying a sequence of control pulses that average out the effects of environmental interactions over time. Dynamically corrected gates \cite{green2012high, cerfontaine2014high} enhance noise robustness by first applying a `naive rotation' followed by a carefully designed, extended identity operation. However, these methods often rely on effectively suppressing or modifying the system-bath interaction, which can make them less flexible or efficient. They may also overlook important environmental factors, such as temperature, non-Markovian effects, or the strength of system-bath coupling.

Traditional quantum control methods often rely on complex mathematical models and extensive experimental data, which present significant limitations in practical applications \cite{Brif2010}. To circumvent this bottleneck, the deep learning methods provide promising strategies for general-purpose optimal control \cite{niu2019universal}, may automatically search the optimal control strategy via study algorithms \cite{chen241319}, and even suppress noise influence in quantum open system \cite{chen25eadr0875}. The gradual progress in this direction has translated control techniques from classical optimal algorithms into artificial intelligence \cite{norambuena2024physics,zhang2025robust}.
Robust and high-precision quantum control can be translated into a Supervised Learning (SL) task by thinking of the time-ordered quantum evolution as a layer-ordered neural network (NN) \cite{wu2019learning}. In addition, using parameters such as the phase, detuning, or Rabi frequency in composite pulse systems as training parameters can achieve robust quantum control \cite{shi2024supervised}. We have also proposed a strategy to optimize quantum control by parameterizing the effects of the environment and incorporating them as features into the training process of the NN, enabling it to generate the optimal control actions for different environments \cite{wang2024adaptive}.
However, the SL algorithm needs to discretize the complete optimization task into single-step tasks when dealing with multi-step optimization tasks. It also requires manually preparing a large amount of labeled data for training. 
Deep reinforcement learning (DRL) algorithms combine the powerful nonlinear modeling capabilities of deep learning with the complex decision-making capabilities of reinforcement learning, enabling them to be applied to a variety of quantum control tasks \cite{he2021deep, zhang2019does, niu2019universal}.
K. Reuer et al.\cite{reuer2023realizing} implemented a real-time quantum feedback control agent based on deep reinforcement learning. By constructing a low-latency neural network on a field-programmable gate array, they successfully initialized a superconducting qubit efficiently, taking a new step in controlling quantum devices with DRL.
Combining the network training strategy we proposed earlier with the advantages of DRL to train a noise-aware model that can generate corresponding optimal control strategies according to different environments is undoubtedly a highly promising and prospective research direction.

In this paper, taking AQSP of a two-level system in a bosonic environment as an example \cite{mikhailov2021non}, we respectively employed the DRL and SL algorithms, incorporating noise into the training process of the neural network to train noise-aware models. The physics systems for this model can be superconducting qubits \cite{potovcnik2018studying}, quantum dots \cite{mi2017strong}, and micromechanical resonators \cite{groeblacher2015observation}.
Our results show that both algorithms are suitable for pulse design in a noisy environment. The DRL algorithm outperforms the SL algorithm in noise-free or weak-noise cases, while the SL algorithm is more effective when environmental effects on the system are stronger. Furthermore, to analyze the computational resources required, we examine the time each algorithm takes to design control trajectories, along with their average step size. The DRL algorithm is more time-consuming than the SL algorithm, especially as noise levels increase.

\section{Model}

\subsection{Open quantum system}

We use the theory of open quantum systems to calculate the evolution of quantum systems in noisy environments. The total Hamiltonian for an open quantum system can be expressed as: $H_{tot}=H_s+H_b+H_{int}$, where $H_s$, $H_b$ and $H_{int}$ represent the Hamiltonian of the system, the bath, and the interaction between the system and the bath, respectively. For a bosonic bath $H_b=\sum_k\omega_kb_k^\dagger b_k$, where $\omega_k$ denotes the frequency of the $k$-th mode of the bath, and $b_k$ ($b_k^\dagger$) is the annihilation (creation) operator, with the commutation relation $[b_{k},b_{k}^{\dagger}]=1$ (where we set the reduced Planck constant $\hbar=1$ from now on). Furthermore, the interaction Hamiltonian $H_{int}$ can be written as: $H_{int}=\sum_k(g_k^*L^\dagger b_k+g_kLb_k^\dagger)$, where $L$ is the Lindblad coupling operator, and $g_k$ is the coupling strength between the system and the $k$-th mode of the bath.

In this study, we utilize the quantum state diffusion (QSD) equation approach, as referenced in \cite{diosi1997non, wang2021quantum}, to analyze the dynamics of the system. The underlying concept involves projecting the total wave function onto the coherent state of the bath, thereby extracting the wave function of the system. To achieve this, an ansatz operator $O$ is introduced, defined by the relation: $\frac{\delta|\psi_{t}\rangle}{\delta z_{t}^{*}}\:=\:O_{z}(t,s,z^{*})\:|\psi_{t}\rangle$, where $z_t$ represents the complex Gaussian noise. 

The non-Markovian master equation, derived from the QSD equation as detailed in \cite{wang2021quantum}, is expressed as:
\begin{equation}
	\begin{aligned}
		\frac{\partial}{\partial t}\rho_{s}=
		&-\:i[H_{s},\rho_{s}]+[L,\rho_{s}\overline{O}_{z}^{\dagger}(t)]-[L^{\dagger},\overline{O}_{z}(t)\rho_{s}]\\
		&+[L^{\dagger},\rho_{s}\overline{O}_{w}^{\dagger}(t)]-[L,\overline{O}_{w}(t)\rho_{s}].
	\end{aligned}
	\label{equ:1}
\end{equation}
where $\overline{O}_{z,(w)}(t)\:=\:\int_{0}^{t}ds\:\alpha_{z,(w)}(t\:-\:s)O_{z,(w)}(t)$, and $\alpha_{z,(w)}(t-s)$ is the correlation function.

For the Lorentz-Drude model, the spectral density is characterized by $J(\omega)\:=\:\frac{\Gamma}{\pi}\:\frac{\omega}{1+(\omega/\gamma)^{2}}$ \cite{wang2010coherent, ritschel2014analytic, meier1999non}. In this expression, $\Gamma$ represents the coupling strength between the system and the bath, $\gamma$ is the  characteristic frequency of the bath, and $T$ denotes the bath temperature. When $\gamma \rightarrow 0$, the system is subjected to colored noise, indicative of a pronounced non-Markovian effect. Conversely, as $\gamma\to\infty$, the system approaches the Markovian limit. The master equation, as denoted by Eq.~(\ref{equ:1}), can be numerically solved with the aid of the following closed equations presented in \cite{wang2021quantum}:

\begin{equation}
	\begin{aligned}
		\frac{\partial\overline{O}_{z}}{\partial t}=
		&\left(\frac{\Gamma T\gamma}{2}-\frac{i\Gamma\gamma^{2}}{2}\right)L-\gamma\overline{O}_{z}\\
		&+\begin{bmatrix}-iH_s-(L^\dagger\overline{O}_z+L\overline{O}_w),\overline{O}_z\end{bmatrix},
	\end{aligned}
	\label{equ:2}
\end{equation}

\begin{equation}
	\begin{aligned}
		\frac{\partial\overline{O}_{w}}{\partial t}=
		&\frac{\Gamma T\gamma}{2}L^{\dagger}-\gamma\overline{O}_{w}\\
		&+\begin{bmatrix}-iH_s-(L^\dagger\overline{O}_z+L\overline{O}_w),\overline{O}_w\end{bmatrix}.
	\end{aligned}
	\label{equ:3}
\end{equation}
The derivation is detailed in Appendix A.

\subsection{Machine learning assisted pulse design}

We use the task of preparing an arbitrary qubit state from an arbitrary initial qubit state as an example of quantum control. This problem is crucial as a first step in quantum computing tasks and contains sufficient complexity to warrant discussion \cite{zhang2022quantum, zhang2019does}. The Hamiltonian of the system is given by:
\begin{equation}
	H_s(t)=J(t)\sigma_z+h\sigma_x,
	\label{equ:4}
\end{equation}
where $\sigma_z$ and $\sigma_x$ represent the Pauli matrices corresponding to the $z$- and $x$-components, respectively. This Hamiltonian can be used to describe quantum control in semiconductor double quantum dots \cite{petta2005coherent, malinowski2017notch, brunner2011two, bluhm2010enhancing}. Specifically, $h$  represents the Zeeman energy splitting between two spins, which is difficult to adjust in physical implementation and is usually considered a fixed value \cite{zhang2019semiconductor}. $ J(t) $ represents the tunable and non-negative exchange coupling \cite{zhang2019semiconductor}, and to avoid disrupting the charge configuration of the semiconductor double quantum dots, $ J(t) $ is set within a certain range. The interaction between the system and its bath is represented by the Lindblad operator $L=\frac{\cos\phi}{2}\sigma_{x}+\frac{\sin\phi}{2}\sigma_{z}$, where $\phi$ is a tunable parameter that specifies the direction of the $x$-axis and the coupling angle on the $X-Z$ plane \cite{purkayastha2020tunable}. Specifically, for a charge qubit in a double quantum dot , $\phi $ could be tuned by adjusting the detuning and hopping parameters \cite{purkayastha2020tunable}. A discrete set of operable parameters $J$ and $\phi$ is used to compose the control pulse action set $\{a_i\}$. The maximum execution time of a quantum state preparation task is $T_{tot}$, evenly divided into small time intervals of $N$ duration $dt = T_{tot}/N$. At each small interval, an action $a_i$ selected from the action set $\{a_i\}$ can be performed. Our goal is to train noise-aware models that can design control trajectories for varying environmental parameters.

We employ the DRL and SL algorithms respectively to accomplish this goal and compare their performance. To demonstrate the effect of noise suppression, we design three strategies for training the model. Case (1): no environment; Case (2): certain environmental parameters; In this case the environmental parameters are integrated into the QSD calculation process, but does not take as features into the training process; Case (3): environmental parameters as encoded as input features into the machine learning training process. The model trained under Case (1) is referred to as the noise-unaware model. The Models trained under Case (2) and Case (3) are collectively referred to as noise-aware models. 

\begin{figure}[htbp]
	\centering{\includegraphics[width=\columnwidth]{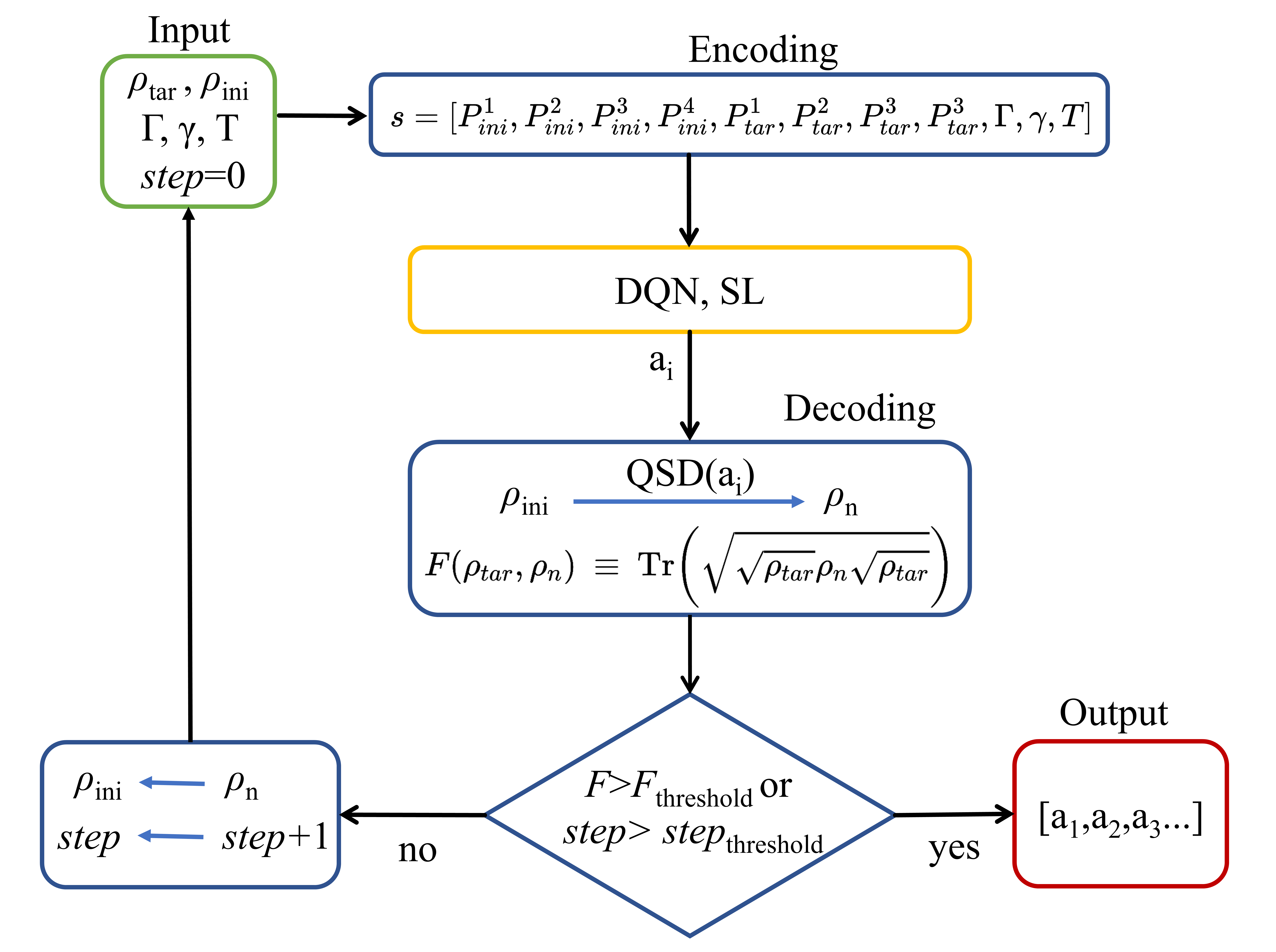}}
	\caption{Design flow chart of control trajectory in Case (3). Input the control task's initial density matrix $\rho_{\text{ini}}$, target density matrix $\rho_{\text{tar}}$, and environmental parameters $\Gamma$, $\gamma$, and $T$. Then encode them into the state $s$. The trained model selects an action $a_i$ based on $s$. Decode and calculate the next quantum state $\rho_n$ and the fidelity $F$. Determine whether the control task is completed based on the preset conditions; if the task is completed, output the control trajectory. If the task is not completed, set $\rho_{\text{ini}} = \rho_n$ and $step = step + 1$, then restart the loop.} 
	\label{fig:1}
\end{figure}

Fig.~\ref{fig:1} shows the process of designing the control trajectory of the trained model in Case (3). First, we encode the initial state $\rho_{\text{ini}}$, target state $\rho_{\text{tar}}$, and environmental parameters $\Gamma$, $\gamma$, and $T$ into a state $s$. Specifically, we use the Positive-Operator Valued Measure (POVM) method to compress $\rho_{\text{ini}}$ and $\rho_{\text{tar}}$ into one-dimensional probability distributions $\{P_{ini}\}$ and $\{P_{tar}\}$. We then encode $\{P_{ini}\}$, $\{P_{tar}\}$, $\Gamma$, $\gamma$, and $T$ into the state $s=[P_{ini}^1, P_{ini}^2,P_{ini}^3,P_{ini}^4, P_{tar}^1,P_{tar}^2,P_{tar}^3,P_{tar}^4, \Gamma, \gamma, T]$ (also known as the input feature sequence). See Appendix B for details on the POVM method. Then, based on $s$, we select an action $a_i$ with the help of a machine learning model while decoding to obtain $\rho_{\text{ini}}$, $\rho_{\text{tar}}$, $\Gamma$, $\gamma$, and $T$. Using the QSD equation method, we calculate the next quantum state $\rho_n$ and the fidelity $F$ between $\rho_n$ and $\rho_{\text{tar}}$. Then determine whether the task is completed by judging the condition $F > F_{\text{threshold}}$ or $step > step_{\text{threshold}}$. If the task is completed, output the control trajectory; otherwise, make $\rho_{\text{ini}} = \rho_n$ enter the loop again. The initial value of $step$ is 0, and $step$ is incremented by 1 with each iteration. If the task cannot be completed even after reaching the maximum number of steps, the model will output the action sequence that achieves the highest fidelity during the intermediate process.

The overall control trajectory design process in Case (1) and Case (2) is the same as in Case (3), but the environmental parameters are not encoded in $s$.

\section{Methods}
\subsection{Deep reinforcement learning}

The DRL algorithm is a method that combines deep learning and reinforcement learning \cite{shalev2014understanding}, allowing an Agent to learn a policy $\pi$ by interacting with an Environment to maximize the cumulative reward $R$. Specifically, the Agent can choose an action $a_i$ based on the state $s$ provided by the Environment. After the Environment executes the action $a_i$, the state $s$ changes to $s'$ and a corresponding reward value $r_t$ is given. Repeating this process until the task is completed or the maximum number of steps is reached allows for the design of a control trajectory. The cumulative reward $R$ for a comprehensive multi-step decision-making task can be expressed as \cite{sutton2018reinforcement}:
\begin{equation}
	R=r_1+\alpha r_2+\alpha^2r_3\cdots+\alpha^{N-1}r_N=\sum_{t=1}^N\alpha^{t-1}r_t,
	\label{equ:5}
\end{equation}
where $\alpha$ represents the discount factor, with values ranging between $[0,1]$, and $N$ is the total number of steps. A larger $R$ indicates better Agent performance. Due to the discount factor, the Agent is naturally inclined to obtain larger rewards in the short term. The objective of the Agent is to find a policy $\pi$ that dictates which action $a_i$ to execute in any given state $s$ to maximize $R$. The action-value function, also known as the $Q$-value \cite{watkins1992q}, is given by:
\begin{equation}
	\begin{aligned}
		Q^\pi(s,a_i)
		&=E[r_t+\alpha r_{t+1}+\cdots|s,a_i]\\
		&=E\Big[r_t+\alpha\:Q^\pi\left(s^{\prime},a^{\prime}\right)|s,a_i\Big].
	\end{aligned}
	\label{equ:6}
\end{equation}
This represents the expected cumulative reward $R$ obtained by executing action $a_i$ in state $s$ under policy $\pi$. In $Q$-learning \cite{watkins1992q}, a $Q$-table is used to record these $Q$-values. With an accurate $Q$-table, the optimal action for a given state $s$ can be easily determined by selecting the action with the highest $Q$-value. 

The essence of the learning process is to continuously update the $Q$-table, with the update formula given by:
\begin{equation}
	Q(s,a_i)\leftarrow Q(s,a_i)+\beta[r_t+\alpha\max_{a'}Q(s',a')-Q(s,a_i)], 
	\label{equ:7}
\end{equation}
where $\beta$ is the learning rate. In updating the $Q$-values, we take into account not only the immediate reward but also the potential future rewards. The revision of the current $Q$-value hinges on the maximum $Q(s',a')$ of the subsequent state $s'$. To determine $\max_{a'}Q(s',a')$, we must compare known action values and explore unknown action values as extensively as possible, which also implies a greater expenditure of resources. Confronted with this trade-off, we employ the $\varepsilon$-greedy algorithm to select actions \cite{he2021deep}. Specifically, we assign a probability $\varepsilon$ to choose the currently most advantageous action and a probability $1-\varepsilon$ to explore unknown actions. As training progresses, $\varepsilon$ gradually increases from $0$ to a value slightly below $1$. This approach enables rapid expansion of the $Q$-table in the initial stages of training and fosters effective $Q$-value updates in the intermediate and later stages of training.

The Deep $Q$-Network (DQN) algorithm \cite{mnih2015human, Mnih2013PlayingAW} employs a multi-layer neural network to supplant the function of a $Q$-table. Specifically, the DQN algorithm utilizes two neural networks with the same architecture: the Main network $\theta$ and the Target network $\theta^-$. These networks are tasked with predicting $Q(s, a_i)$ and $\max_{a'} Q(s', a')$ in Eq.~(\ref{equ:7}), respectively. Updating the $Q$-table thus translates into updating the parameters of these two networks. We employ an experience replay strategy \cite{Mnih2013PlayingAW} to train the Main network $\theta$. Throughout the training process, the Agent accumulates experience unit $(s, a, r, s')$ and stores them in an Experience Memory $D$ with a memory capacity of $M$. The Agent then randomly selects a batch of $N_{bs}$ experience units from the Experience Memory $D$ to train the Main network $\theta$. We use the loss function:
\begin{equation}
	Loss=\frac{1}{N_{bs}}\sum_{i=1}^{N_{bs}}\left(\left[r+\alpha\max_{a'}Q\big(s',a'\big)\right]_i-Q(s,a)_i\right)^2
	\label{equ:8}
\end{equation}
to calculate the loss and employ mini-batch gradient descent (MBGD) algorithm \cite{mnih2015human, Mnih2013PlayingAW} to optimize the parameters of the Main network $\theta$. The Target network $\theta^-$ remains inactive during the training process and only updates its parameters by directly copying from the Main network $\theta$ every $C$ steps.

\subsection{Supervised learning}

In the SL algorithm, the model learns the mapping between input features and output labels by analyzing labeled data sets \cite{cunningham2008supervised, singh2016review,jiang2020supervised, nielsen2015neural}. We denote input features as $s$ and output labels as action $a_i$. The trained model can provide the correct $a_i$ based on the input $s$, perform the action $a_i$ to obtain the next state $s'$, and then update $s \leftarrow s'$. Repeating this process until the completion of preset conditions allows for the design of a control trajectory. The action selection at each step can be considered a multi-classification task, with the number of possible actions $\{a_i\}$. 

Whether the model can choose the correct action given $s$ depends on the appropriateness of the labels when constructing the training data set. A commonly used strategy is the standard greedy algorithm \cite{cormen2022introduction, balaman2019chapter}. Specifically, for multi-step decision-making tasks, each step selects and executes the action that yields the best immediate effect (maximum fidelity). However, the locally optimal action selected by such a strategy may not be globally optimal \cite{lo2017solving, Christof2024}. To mitigate the adverse effects of local optimality, we perform fidelity evaluations at progressive intervals to systematically remove instances with locally optimal actions\cite{li2023enhanced}. Specifically, any data that fails to show enhanced fidelity compared to previous data is considered a symptom of local optimality and is excluded from the data set by us. The model trained under this strategy has a strong ability to avoid local optimality, making its predicted control trajectory better able to perform quantum control tasks \cite{wang2024adaptive}.

We collect each step of each task in the training task set as an experience unit $(s, a)$ to form a training data set. The training data set is randomly shuffled and divided into $80\%$ and $20\%$ parts, with the large part used for training and the smaller part used to verify the performance of the model during training. Negative Log Likelihood Loss (NLLLoss) \cite{yao2020negative} was used as the loss function, and Logarithmic Softmax activation function was applied to the output layer of the model. By minimizing the NLLLoss, which measures the difference between the probability distribution predicted by the model and the true label, the model learns more accurate class predictions. The formula for the loss function is:
\begin{equation}
	Loss=-\frac1{N_{bs}}\sum_{i=1}^{N_{bs}}\log(p_{y_i})
	\label{equ:9}
\end{equation}
where $p_{y_i}$ is the predicted probability of the model for the $i-$th sample to belong to its true class $y_i$, which is obtained by the Logarithmic Softmax activation function. A trained model is able to give the correct class label based on the input $s$, which is the correct control action $a_i$. Through the process shown in Fig.~\ref{fig:1}, we can obtain a complete control trajectory and achieve high precision control in the quantum control task.

The implementation of both the DRL and SL algorithms uses Python 3.8.19 and Pytorch 2.4.1 \cite{paszke2019pytorch}, and runs memory on a 64-core 3.40 GHz 125.6 GB CPU.

\section{Results and discussion}
We sample $32$ quantum states on the Bloch sphere and select different pairs of quantum states as the initial state $\rho_{\text{ini}}$ and the target state $\rho_{\text{tar}}$ for the quantum state preparation task. This forms a task set comprising a total of $992$ $(32 \times 31)$ preparation tasks. After randomly shuffling the task set, we construct a training set with $700$ tasks, a validation set with $50$ tasks, and a test set with $242$ tasks. We set the maximum operation time for a preparation task to $T_{tot}=2\pi$, with the execution time of each step being $\pi/5$, and the maximum number of operation steps $N=10$. We define two discrete sets of controllable parameters: $J\in\{0,1,2,3,4,5,6,7,8\}$ and $\phi\in\{\pi,\:\frac{\pi}{2},\:\frac{\pi}{4}\}$, which correspond to a total of $27$ control action set $\{a_i\}$.

\begin{figure}
	\centering{\includegraphics[width=\columnwidth]{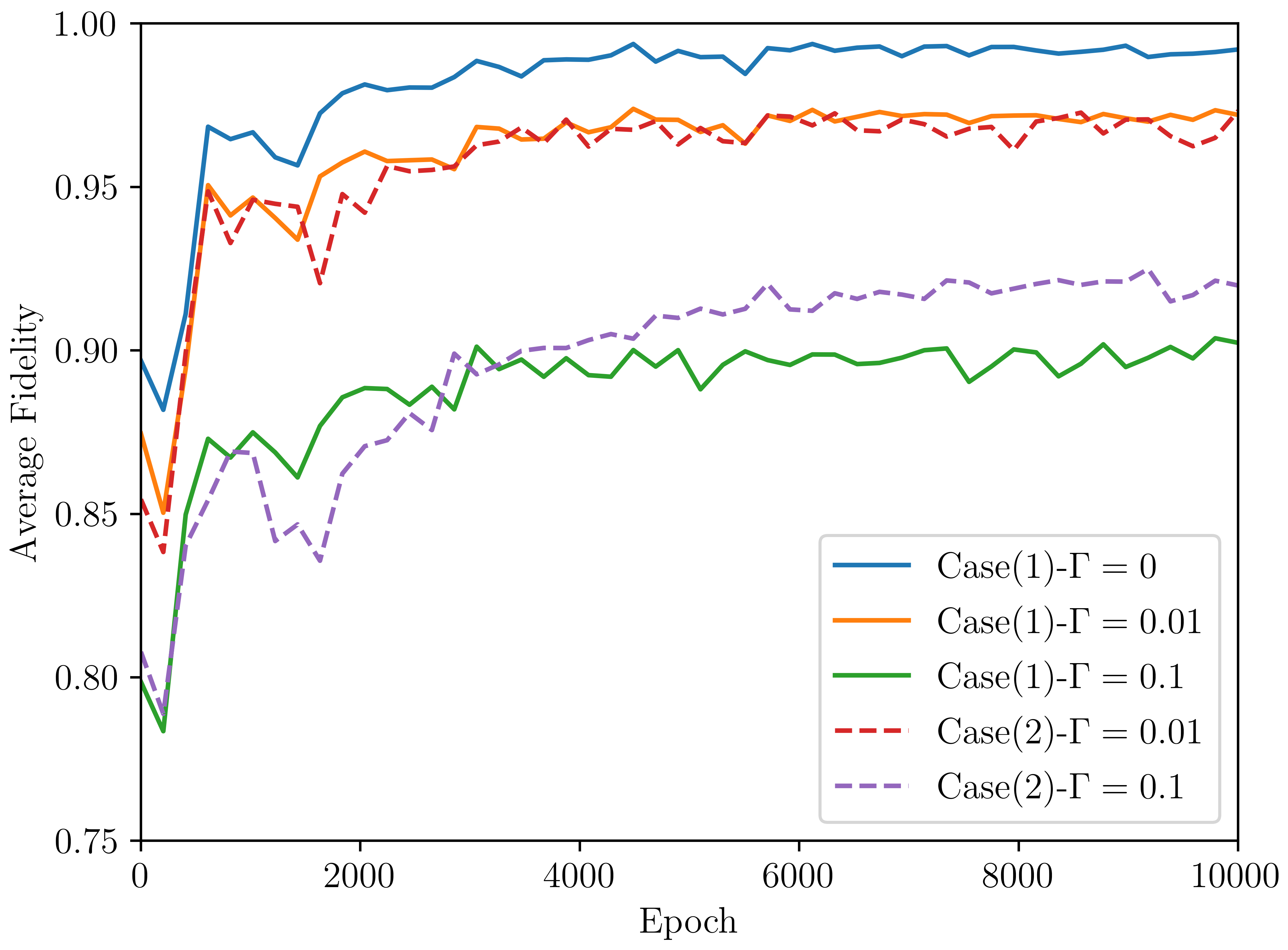}}
	\caption{The validation set average fidelity $\bar{F}$ of the model trained by the DRL algorithm as a function of training epoch under Case (1) and Case (2). The maximum training epoch is 10000, and the model effect is verified on the validation set every 200 epoch.}
	\label{fig:2}
\end{figure}

First, we discuss the training process using the DRL algorithm in Case (1) and Case (2). As shown in Fig.~\ref{fig:2}, the average fidelity $\bar{F}$ of all models reaches a stable value after an initial period of rapid growth, indicating that the models have completed training. The three solid lines represent the verification results of the model in three different environments during training under the Case(1) strategy. When there is no noise environment $(\Gamma=0)$ and a weak noise $(\Gamma=0.01)$, the performance of the model is good, but when the noise intensity increases, the performance of the model drops sharply. To better suppress noise, we adopt the strategy of Case (2) to train the model for a specific noise environment. When the weak noise is present $(\Gamma=0.01)$, the performance of the models trained by the two strategies is similar, but when the noise becomes stronger $(\Gamma=0.1)$, the model trained by Case (2) strategy exhibits superior noise resistance. Note that the SL algorithms have the same conclusion \cite{wang2024adaptive}.

\begin{figure}
	\centering{\includegraphics[width=\columnwidth]{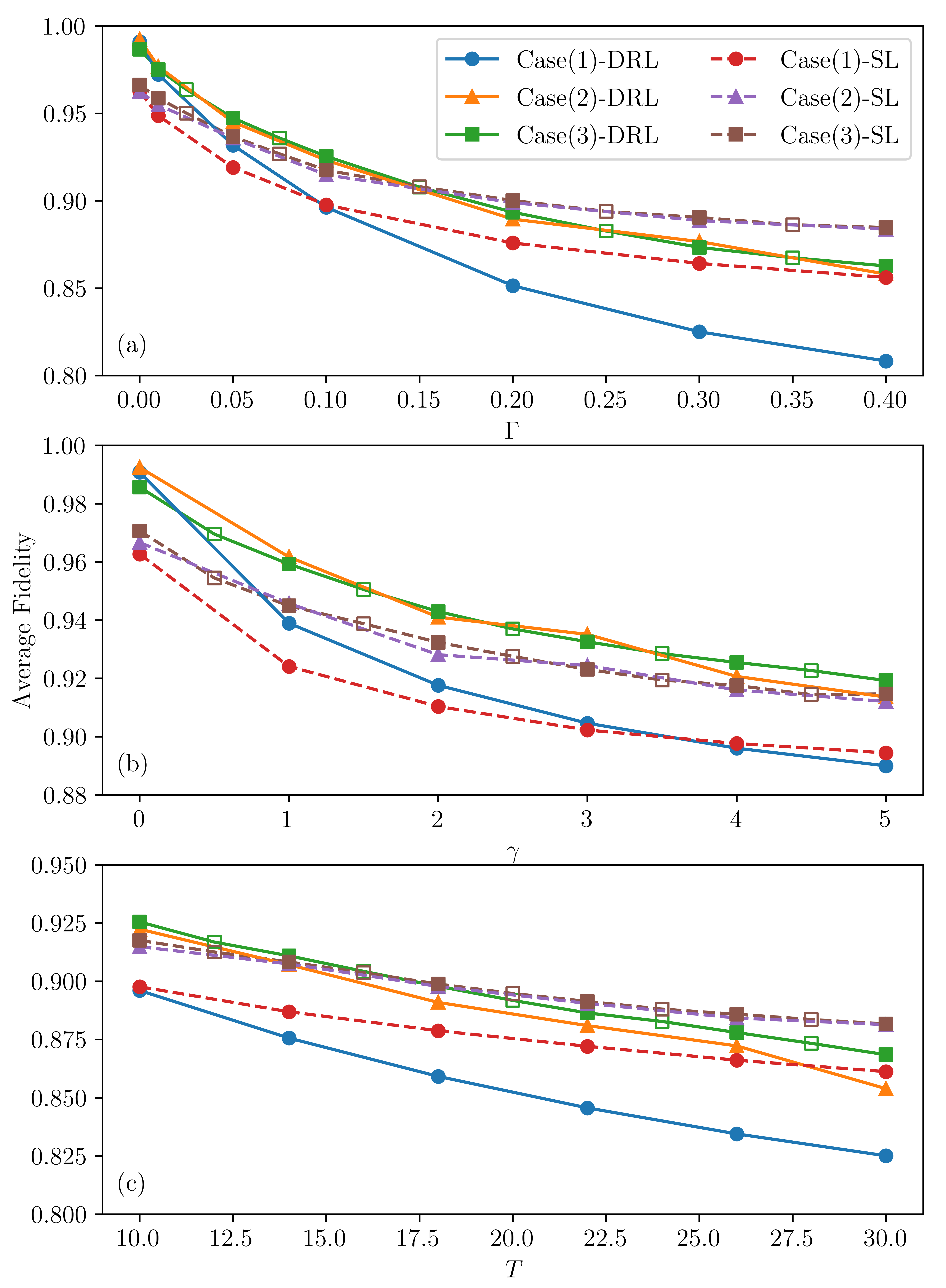}}
	\caption{The test set average fidelity $\bar{F}$ of the model trained by the DRL and SL algorithms as a function of different environmental parameters: (a) $ \Gamma$ $(\gamma = 4, T = 10)$; (b) $\gamma$  $(\Gamma = 0.1, T = 10)$; and (c) $ T$  $(\Gamma = 0.1, \gamma = 4)$ for three cases. The hollow data points represent the test results of environmental parameters that the model has not seen during the training process.}
	\label{fig:3}
\end{figure}

Next, we discuss the performance of models trained by the DRL and SL algorithms under different Cases. Given the distinct training logic of these two algorithms, we focus solely on the performance of the convergent models trained by both algorithms with the same neural network architecture. For Case (1), we train a model that disregards the environment and test it under various environmental parameters. In Case (2), we train a model for each set of environmental parameters and test it against the corresponding parameters. Case (3) incorporates environmental parameters as features into the machine learning training process, enabling the trained model to self-adapt to different environment, and we only train one model and test it under different environmental parameters. To determine whether the trained models have converged, we use the average cumulative reward and the average loss value from the validation set during the training process as criteria for both the DRL and SL algorithms. The training process and hyperparameter table of each model of the two algorithms in different cases are shown in Appendix C.	

Fig.~\ref{fig:3} (a-c) plots the test set average fidelity $\bar{F}$ as a function of different environmental parameters $\Gamma$  $(\gamma = 4, T = 10)$, $\gamma$ $(\Gamma = 0.1, T = 10)$ and $T$ $(\Gamma = 0.1, \gamma = 4)$ for the two algorithms under three Cases. From Fig.~\ref{fig:3}, first, $\bar{F}$ decreases with increasing $\Gamma$, $\gamma$ and $T$ in all cases for both algorithms. This indicates a stronger interaction intensity, a more Markovian environment, and a higher temperature will destroy more of the quantumness of the system. Thus, they correspond to a lower average fidelity $\bar{F}$. These results are in accordance with our previous observation \cite{wang2024adaptive,wang2021quantum}. Secondly, the performance of Case (2) and (3) is better than that of Case (1), this means that the network with consideration of environmental parameters is more powerful. Both algorithms show that the difference in $\bar{F}$ between Case (1) and Cases (2) (3) is relatively small for a weak environment. This difference becomes bigger with increasing $\Gamma$, $\gamma$ and $T$, indicating a stronger environment provides more space for the NN to enhance $\bar{F}$ through optimization. Thirdly, similar values of $\bar{F}$ are obtained for Cases (2) and (3) using the same algorithm. This demonstrates the superiority of Case (3), i.e., by taking the environmental parameters as input features, a well-trained NN has the ability to design optimal pulses for different environment parameters, even for varying parameters. We emphasize that this algorithm is general: once the environmental parameters are given and incorporated into the training, the NN can deal with complex environment noise. At last, comparing the performance of the two algorithms in the same case, it is interesting to see that when there is no noise ($\Gamma=0$) or the noise is relatively weak, the DRL algorithm outperforms the SL algorithm in all three cases. However, the SL algorithm becomes more advantageous in strongly noisy environments. The DRL algorithm can handle continuous decision-making processes \cite{shalev2014understanding}, which is crucial for the design of multi-step control trajectories. In contrast, when the SL algorithm is applied to multi-step control problems, it makes choices step by step, and it is challenging to ensure that each local choice is the wisest in the overall context. During the process of model training, the DRL algorithm can realize end-to-end learning and design control trajectory directly from the initial state to the target state, reducing the dependence on intermediate steps. In contrast, the SL algorithm needs to split the complete multi-step control trajectory into single-step labeled data to train the model. 
For these reasons, the DRL algorithm can design more accurate and effective control strategies than the SL algorithm in the noise-free or noise-weak conditions. However, under the influence of strong noise, the advantages of the DRL algorithm and the disadvantages of the SL algorithm are both mitigated by the noise impact. Specifically, noise reduces the reward $r$ of the Agent's chosen action $a_i$ in the DRL algorithm, leading to a deterioration in its interaction with the Environment. In contrast, the stronger stability of the SL algorithm makes it more suitable for dealing with strong noise.

\begin{figure}[htbp]
	\centering{\includegraphics[width=\columnwidth]{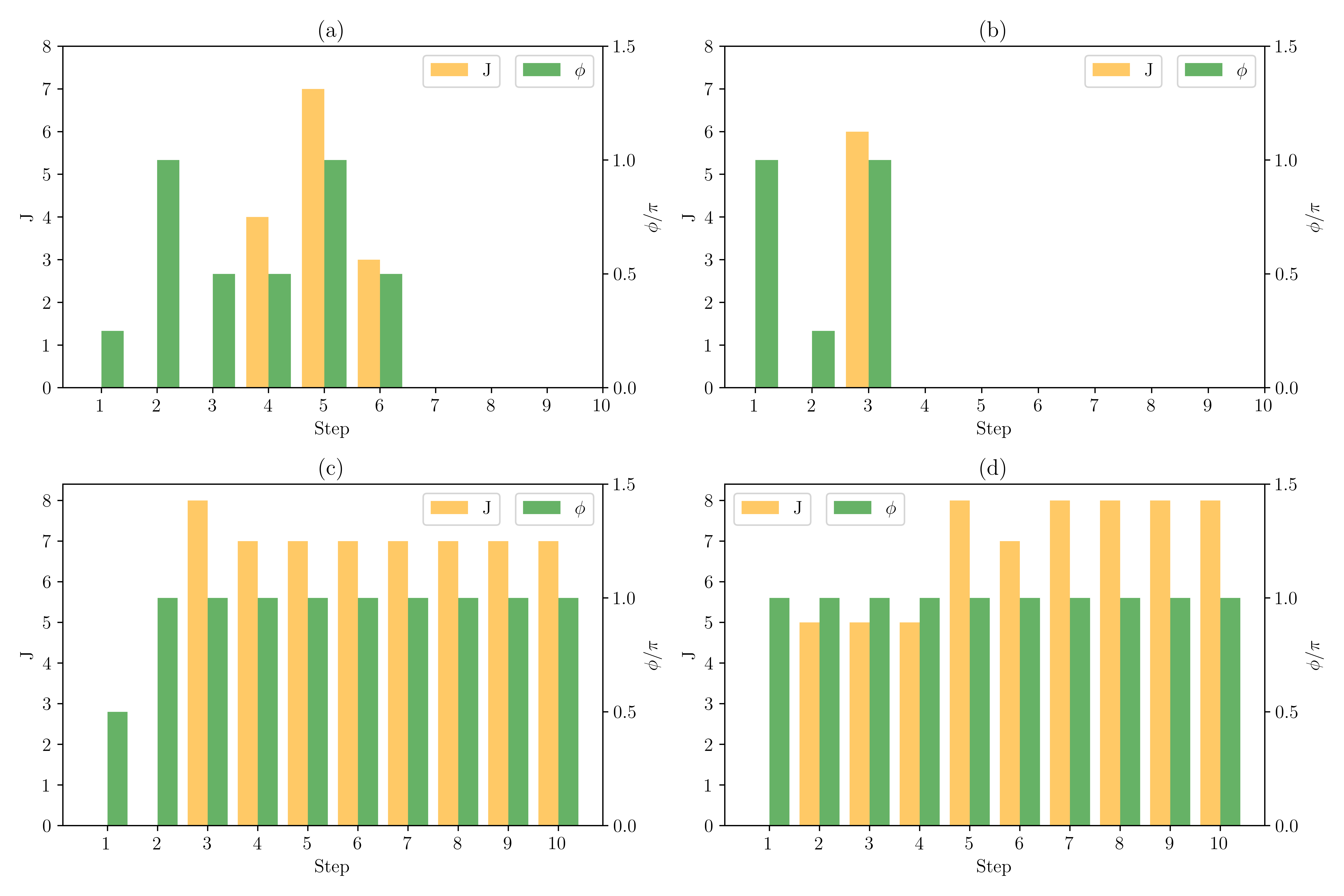}}
	\caption{Comparison of the control action sequences generated by the DRL and SL models trained in Case (3) under different environmental parameters. The initial and target states are set to $ |0\rangle $ and  $ |1\rangle $. (a) and (c) show the control action sequences generated by DRL for $ \Gamma = 0.01 $ and $ \Gamma = 0.4 $, respectively. (b) and (d) show the control action sequences generated by SL for $ \Gamma = 0.01 $ and $ \Gamma = 0.4 $, respectively. $ \gamma = 4 $, $ T = 10$ .} 
	\label{fig:4}
\end{figure}

To more intuitively illustrate the pulse control strategies designed by the two algorithms, we have plotted the pulse action sequences for preparing the state $ |1\rangle $ from the state $ |0\rangle $ in Fig.~\ref{fig:4}. The action sequences designed by the two algorithms for the same task in the same environment are not the same. For  $ \Gamma = 0.01 $ , the fidelities achieved by DRL and SL after the final step are $ 0.9759 $ and $0.9545$, respectively. For $ \Gamma = 0.4 $, the results are $0.8464$ and $0.8572$.

\begin{figure}
	\centering{\includegraphics[width=\columnwidth]{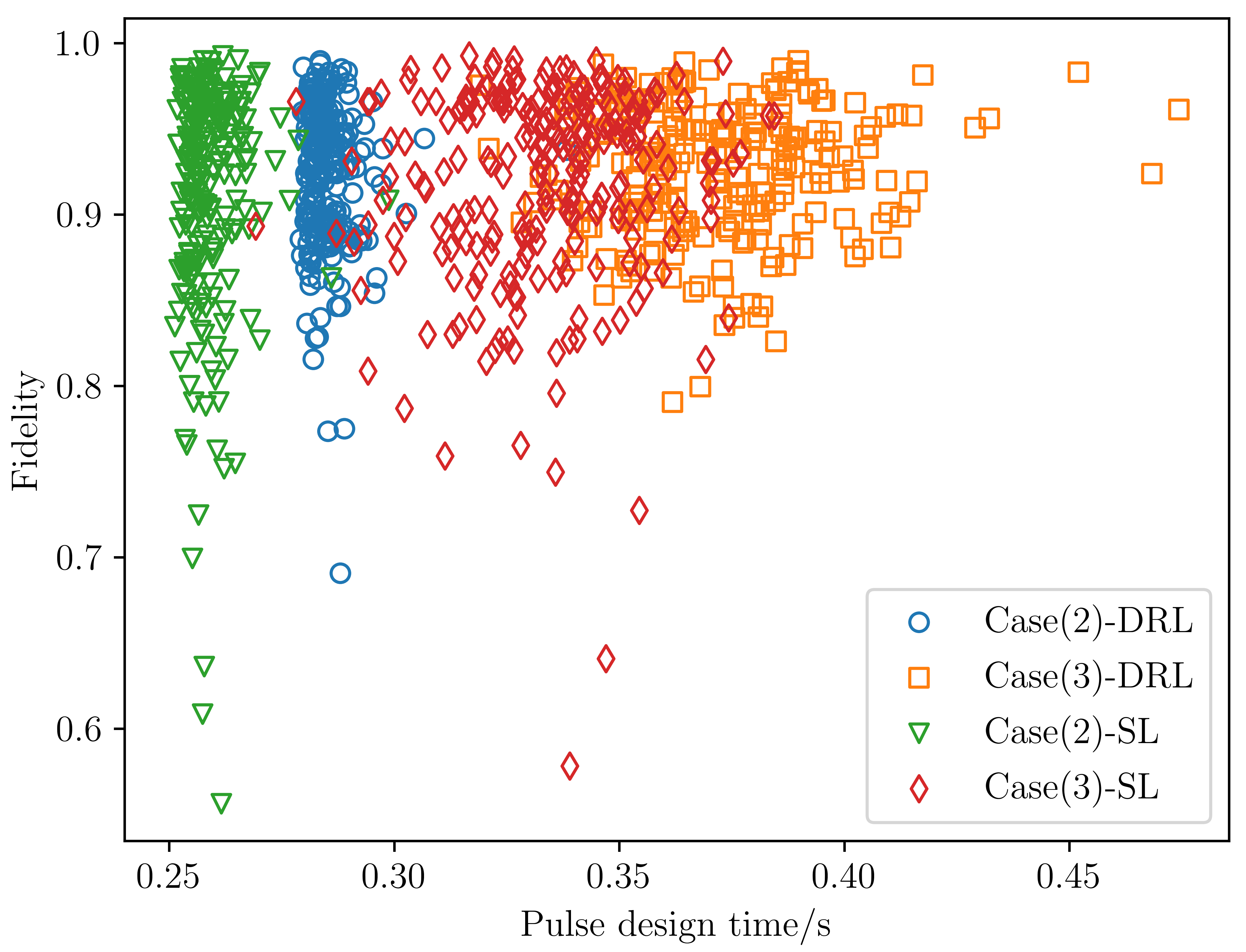}}
	\caption{The distribution of the fidelities $F$ versus design time of the DRL and SL algorithms in Case (2) and Case (3) based on the test set. $\Gamma=0.1$,$\gamma=4$ and $T=10$.}
	\label{fig:5}
\end{figure}	

Finally, we compare the time cost and the step size taken by the DRL and SL algorithms to design control trajectories. The environmental parameters are set as $\Gamma=0.1$, $\gamma=4$ and $T=10$. Fig.~\ref{fig:5} is a scatterplot of pulse design time and fidelity for the test set tasks. Tab.~\ref{tab:1} shows the average fidelity $\bar{F}$, the average design time $\bar{t}$ and the average design action step size $\bar{n}$ of the test set under different conditions. Models trained by both algorithms in Case (2) require less design time than those trained in Case (3) due to the simpler neural network structure. However, the advantage of Case (3) is that the well trained NN is suitable for arbitrary environmental parameters.  
The design time distribution of the two models trained by the SL algorithm is concentrated around $0.2588$ and $0.3334$ seconds, while that for the DRL algorithm is more dispersed. This indicates that the SL algorithm is more stable than the DRL algorithm.  Within the same case, the DRL algorithm, with its more complex interaction logic, takes more time than the SL algorithm. 

\begin{table}[htbp]
	\centering
	\caption{The average fidelity $\bar{F}$, average design time $\bar{t}$ and average design action step size $\bar{n}$ of the DRL and SL algorithms in Case(2) and Case(3) based on the test set, respectively.}
	\renewcommand{\arraystretch}{1.5} 
	\begin{tabular*}{\columnwidth}{@{\extracolsep{\fill}}cccc}
		\hline
		Conditions     & $\bar{F}$  & $\bar{t}$ & $\bar{n}$ \\
		\hline
		Case(2)-DRL    & $ {0.9232}^{+0.0665}_{-0.2325} $   & $ {0.2850}^{+0.0531}_{-0.0059} $& 2.8430 \\
		Case(3)-DRL    & $ {0.9255}^{+0.0642}_{-0.1349} $   & $ {0.3725}^{+0.1018}_{-0.0533} $& 3.0537 \\
		Case(2)-SL     & $ {0.9148}^{+0.0778}_{-0.3586} $   & $ {0.2588}^{+0.0399}_{-0.0076} $& 2.5579 \\
		Case(3)-SL     & $ {0.9175}^{+0.0751}_{-0.3393} $   & $ {0.3334}^{+0.0510}_{-0.0642} $& 2.4917 \\
		\hline
	\end{tabular*}
	\label{tab:1}
\end{table}

\section{Conclusions}

We use the DRL and SL algorithms, respectively, to train models capable of designing pulses under different noise conditions. When comparing the performance of the two algorithms, we find that the DRL algorithm, with its powerful dynamic decision-making and end-to-end learning capabilities, outperforms the SL algorithm, especially in noise-free or low-noise conditions. In contrast, the SL algorithm is more stable, with models trained using it remaining accurate even as environmental disturbances increase.

Furthermore, incorporating environmental parameters as input features into model training offers a resource-efficient solution. It minimizes the need to customize multiple models for different environments, reducing overall computational costs, though at the expense of requiring more neurons and longer pulse design times. This approach enables universal control strategies applicable across a range of noise intensities.

\section*{Acknowledgments}
We acknowledge Tuo-Zhi Chen for his valuable input and thoughtful discussions while this manuscript was being prepared. 
This paper is supported by the Natural Science Foundation of Shandong Province (Grant No.ZR2024MA046, ZR2021LLZ004) and Fundamental Research Funds for the Central Universities (Grant No.202364008). 
Project for PhD Students.Y.-A.Y. is supported by the National Natural Science Foundation of China (NSFC) under Grant No. 21973036. L.-A.W. is supported by the Basque Country Government (Grant No.IT1470-22) and Grant No. PGC2018-101355-B-I00 funded by MCIN/AEI/10.13039/501100011033, This project has also received support from the Spanish Ministry for Digital Transformation and of Civil Service of the Spanish Government through the QUANTUM ENIA project call - Quantum Spain, EU through the Recovery, Transformation and Resilience Plan-NextGenerationEU within the framework of the Digital Spain 2026.

\appendix
\section{Derivations of the non-Markovian master equation}
\label{app:derivation}

In this supplementary section, we detail the derivation of the non-Markovian master equation. The wave function of the system is derived by projecting the coherent states of the bath onto the wave function of the total system. The pure state of the system evolves in accordance with \cite{strunz1999quantum, yu2004non}:	
\begin{equation}
	\begin{aligned}
		\frac{\partial}{\partial t}\left|\psi(t,z_{t}^{*},w_{t}^{*})\right\rangle\:
		&=\:[-iH_{s}+Lz_{t}^{*}+L^{\dagger}w_{t}^{*}-L^{\dagger}\overline{O}(t,z_{t}^{*},w_{t}^{*})\\
		&-L\overline{Q}(t,z_{t}^{*},w_{t}^{*})]\left|\psi(t,z_{t}^{*},w_{t}^{*})\right\rangle,
	\end{aligned}
	\label{equ:a1}
\end{equation}
where $z_{t}^{*}$ and $w_{t}^{*}$ are complex Gaussian noise of environment. The operators $\overline{O}$ and $\overline{Q}$, which account for the environmental memory effects, are defined as:
\begin{equation}
	\overline{O}(t,z_t^*,w_t^*)\:=\:\int_0^tds\alpha(t,s)O(t,s,z_t^*,w_t^*),
	\label{equ:a2}
\end{equation}
\begin{equation}
	\overline{Q}(t,z_t^*,w_t^*)\:=\:\int_0^tds\eta(t,s)Q(t,s,z_t^*,w_t^*),
	\label{equ:a3}
\end{equation}
where $\alpha(t,s)$ and $\eta(t,s)$ are the bath correlation functions, given by: 
\begin{equation}
	\alpha(t,s)=\int d\omega J(\omega)(\overline{n}_k+1)e^{-i\omega_k(t-s)},
	\label{equ:a4}
\end{equation}
\begin{equation}
	\eta(t,s)=\int d\omega J(\omega)\overline{n}_ke^{i\omega_k(t-s)}.
	\label{equ:a5}
\end{equation}
Here, $\overline{n}_{k}\:=\:\frac{1}{e^{\omega_{k}/T}-1}$ is the average number of thermally occupied quanta in the mode $\omega_{k}$, and $J(\omega)$ is the bath spectral density \cite{de2017dynamics}.

The functional derivatives ansatz of the operators $\overline{O}$ and $\overline{Q}$ are given by:
\begin{equation}
	O(t,s,z_t^*,w_t^*)\left|\psi(t,z_t^*,w_t^*)\right\rangle=\frac{\delta}{\delta z_s^*}\left|\psi(t,z_t^*,w_t^*)\right\rangle,
	\label{equ:a6}
\end{equation}
\begin{equation}
	Q(t,s,z_t^*,w_t^*)\left|\psi(t,z_t^*,w_t^*)\right\rangle=\frac{\delta}{\delta w_s^*}\left|\psi(t,z_t^*,w_t^*)\right\rangle.
	\label{equ:a7}
\end{equation}
The $O$ and $Q$ operators describe the evolution of the state at time $t$ as influenced by its past $s$ dependence on the noises $z_{t}^{*}$ and $w_{t}^{*}$.

From the conditions of consistency, we have:
\begin{equation}
	\begin{aligned}
		\frac{\partial O_{z}}{\partial t}=
		&\left[-\mathrm{i}H_{s}+Lz_{t}^{*}-L^{\dagger}\overline{O}_{z}+L^{\dagger}w_{t}^{*}-L\overline{O}_{w},O_{z}\right]\\
		&-\left(L^{\dagger}\frac{\delta\overline{O}_{z}}{\delta z_{s}^{*}}+L\frac{\delta\overline{O}_{w}}{\delta z_{s}^{*}}\right),
	\end{aligned}
	\label{equ:a8}
\end{equation}
\begin{equation}
	\begin{aligned}
		\frac{\partial O_{w}}{\partial t}=
		&\left[-\mathrm{i}H_{s}+Lz_{t}^{*}-L^{\dagger}\overline{O}_{w}+L^{\dagger}w_{t}^{*}-L\overline{O}_{w},O_{z}\right]\\
		&-\left(L^{\dagger}\frac{\delta\overline{O}_{z}}{\delta z_{s}^{*}}+L\frac{\delta\overline{O}_{w}}{\delta z_{s}^{*}}\right).
	\end{aligned}
	\label{equ:a9}
\end{equation}

The reduced density operator of the system is given by the ensemble average of numerous quantum trajectory realizations:
\begin{equation}
	\rho_s = M[P_t]
	\label{equ:a10}
\end{equation}
where $P_{t}\:=\:|\psi(t,z_{t}^{*},w_{t}^{*})\rangle\:\langle\psi(t,z_{t}^{*},w_{t}^{*})|$, and $M[\cdot]$ denotes the ensemble average over the complex Gaussian noise $z_{t}^{*}$ or $w_{t}^{*}$, and $M[\mathcal{F}]=\prod_{k}\frac{1}{\pi}\int e^{-|z|^{2}}\mathcal{F}d^{2}z$. Taking the time derivative of $\rho_s$, we arrive at the non-Markovian master equation \cite{diosi1998non}:
\begin{equation}
	\begin{aligned}
		\frac{\partial}{\partial t}\rho_{s}=
		&-i\left[H_{s},\rho_{s}\right]+[L,M[P_{t}\overline{O}^{\dagger}(t,z_{t}^{*},w_{t}^{*})]]\\
		&-[L^{\dagger},M[\overline{O}(t,z_{t}^{*},w_{t}^{*})P_{t}]]\\
		&+[L^{\dagger},M[P_{t}\overline{Q}^{\dagger}(t,z_{t}^{*},w_{t}^{*})]]\\
		&-[L,M[\overline{Q}(t,z_{t}^{*},w_{t}^{*})P_{t}]].
	\end{aligned}
	\label{equ:a11}
\end{equation}
For the case of weak system-bath coupling, the noise-dependent operators $\overline{O}^{\dagger}(t,z_{t}^{*},w_{t}^{*})$ and $\overline{Q}^{\dagger}(t,z_{t}^{*},w_{t}^{*})$ can be approximated by the noise-independent operators $\overline{O}^{\dagger}(t)$ and $\overline{Q}^{\dagger}(t)$. Thus, the non-Markovian master equation, Eq.~(\ref{equ:1}), is derived.

In the context of the Lorentz-Drude spectrum, the correlation functions depicted in Eqs.~(\ref{equ:a2}-\ref{equ:a3}) are governed by the following equation:
\begin{equation}
	\frac{\partial\alpha_{z(w)}(t-s)}{\partial t}=-\gamma_j\alpha_{z(w)}(t-s).
	\label{equ:a12}
\end{equation}
Leveraging these relationships along with Eqs.~(\ref{equ:a8}-\ref{equ:a9}), we are able to derive Eqs.~(\ref{equ:2}-\ref{equ:3}). Armed with these derived equations, the solution to the Eq.~(\ref{equ:1}) can be pursued numerically.

\section{Positive-operator valued measure}
\label{app:povm}

The POVM method can transform a density matrix containing complex numbers into a probability distribution containing only real numbers. This is particularly useful for addressing the convergence of quantum information and machine learning \cite{carrasquilla2021probabilistic, reh2021time, carrasquilla2019reconstructing, luo2022autoregressive}. Specifically, a set of positive semi-definite measurement operators $\mathbf{M}=\{M_{(\mathbf{a})}\}$ is used to map the density matrix onto a corresponding set of measurement outcomes $\mathbf{P}=\{P_{(\mathbf{a})}\}$, where the vector $\mathbf{a}$ denotes the sequence associated with various measurement bases. When these outcomes fully represent the information content of the density matrix, they are known as informationally complete POVMs (IC-POVMs). These operators satisfy the normalization condition $\sum_{\mathbf{a}}M_{(\mathbf{a})}=\mathbb{I}$.

For an $N$-qubit system, the density matrix can be transformed into a probability distribution by: 
\begin{equation}
	P_{(\mathbf{a})}=\operatorname{tr}[\rho M_{(\mathbf{a})}],
	\label{equ:b1}
\end{equation}
where $M_{(\mathbf{a})}=M_{(a_{1})}\otimes...\otimes M_{(a_{N})}$. By inverting Eq.~(\ref{equ:b1}), the density matrix $\rho$ can be reconstructed as:
\begin{equation}
	\rho=\sum_\mathbf{a}\sum_{\mathbf{a}'}P(\mathbf{a})T_{\mathbf{a}a'}^{-1}M(\mathbf{a}')
	\label{equ:b2}
\end{equation}
where $T_{\mathbf{a}\mathbf{a}^{\prime}}=\mathrm{tr}(M_{(\mathbf{a})}M_{(\mathbf{a}^{\prime})})$ represents an element of the overlap matrix $T$. In this work, we use the Pauli-4 POVM $\mathbf{M}_{\text{Pauli}-4}$\cite{carrasquilla2019reconstructing}, where each sub-measurement operator $M_{(a_{i})}$ is selected from $\{M^{1}=\frac{1}{3}\times|0\rangle\langle0|,M^{2}=\frac{1}{3}\times|l\rangle\langle l|,M^{3}=\frac{1}{3}\times|+\rangle\langle+|,M^{4}\:=\mathbb{I}-M^1-M^2-M^3\}$.

\section*{APPENDIX C: The training process of the models and the hyperparameter tables}

\setcounter{equation}{0}	
\renewcommand\theequation{C.\arabic{equation}}

\begin{table}[htbp]
	\centering
	\caption{Hyperparameter table of the DRL algorithm}
	\renewcommand{\arraystretch}{1.5}
	\begin{tabular*}{\columnwidth}{@{\extracolsep{\fill}}ccc}
		\hline
		Parameters & Case (1) or (2) & Case (3) \\
		\hline
		Total time $T$  & $2\pi$ & $2\pi$    \\
		Action duration $\mathrm{d}t$  & $\pi/5$ & $\pi/5$    \\
		Maximum steps $N$  & $10$ & $10$    \\
		Batch size $N_{bs}$ & $32$ & $32$    \\
		Learning rate $\alpha $ & $0.002$ & $0.001$    \\
		Neurons per hidden layer & $32/64/32$ & $44/88/176/88/44$ \\
		Fidelity threshold $F_{threshold}$ & $0.999$ & $0.999$    \\
		Epoch for training  & $10000$ & $10000$    \\
		Activation function & Relu & Relu    \\
		Memory size $M$ & $20000$ & $20000$    \\
		Replace period $C$ & $200$ & $200$    \\
		Reward discount factor $\gamma$ & $0.9$ & $0.9$    \\
		$\epsilon$-greedy increment $\delta\epsilon$  & $0.001$ & $0.001$    \\
		Maximal $\epsilon$ in training $\epsilon_{max}$  & $0.95$ & $0.95$    \\
		$\epsilon$ in validation and test  & $1$ & $1$    \\
		\hline
	\end{tabular*}
	\label{tab:2}
\end{table}	

\begin{table}[htbp]
	\centering
	\caption{Hyperparameter table of the SL algorithm}
	\renewcommand{\arraystretch}{1.5}
	\begin{tabular*}{\columnwidth}{@{\extracolsep{\fill}}ccc}
		\hline
		Parameters & Case (1) or (2) & Case (3) \\
		\hline
		Total time $T$  & $2\pi$ & $2\pi$    \\
		Action duration $\mathrm{d}t$  & $\pi/5$ & $\pi/5$    \\
		Maximum steps $N$  & $10$ & $10$    \\
		Batch size $N_{bs}$ & $16$ & $16$    \\
		Learning rate $\alpha $ & $0.001$ & $0.001$    \\
		Neurons per hidden layer & $32/64/32$ & $44/88/176/88/44$ \\
		Fidelity threshold $F_{threshold}$ & $0.999$ & $0.999$    \\
		Epoch for training  & $500$ & $500$    \\
		Activation function & Relu & Relu    \\
		\hline
	\end{tabular*}
	\label{tab:3}
\end{table}	

\begin{figure}
	\centering{\includegraphics[width=\columnwidth]{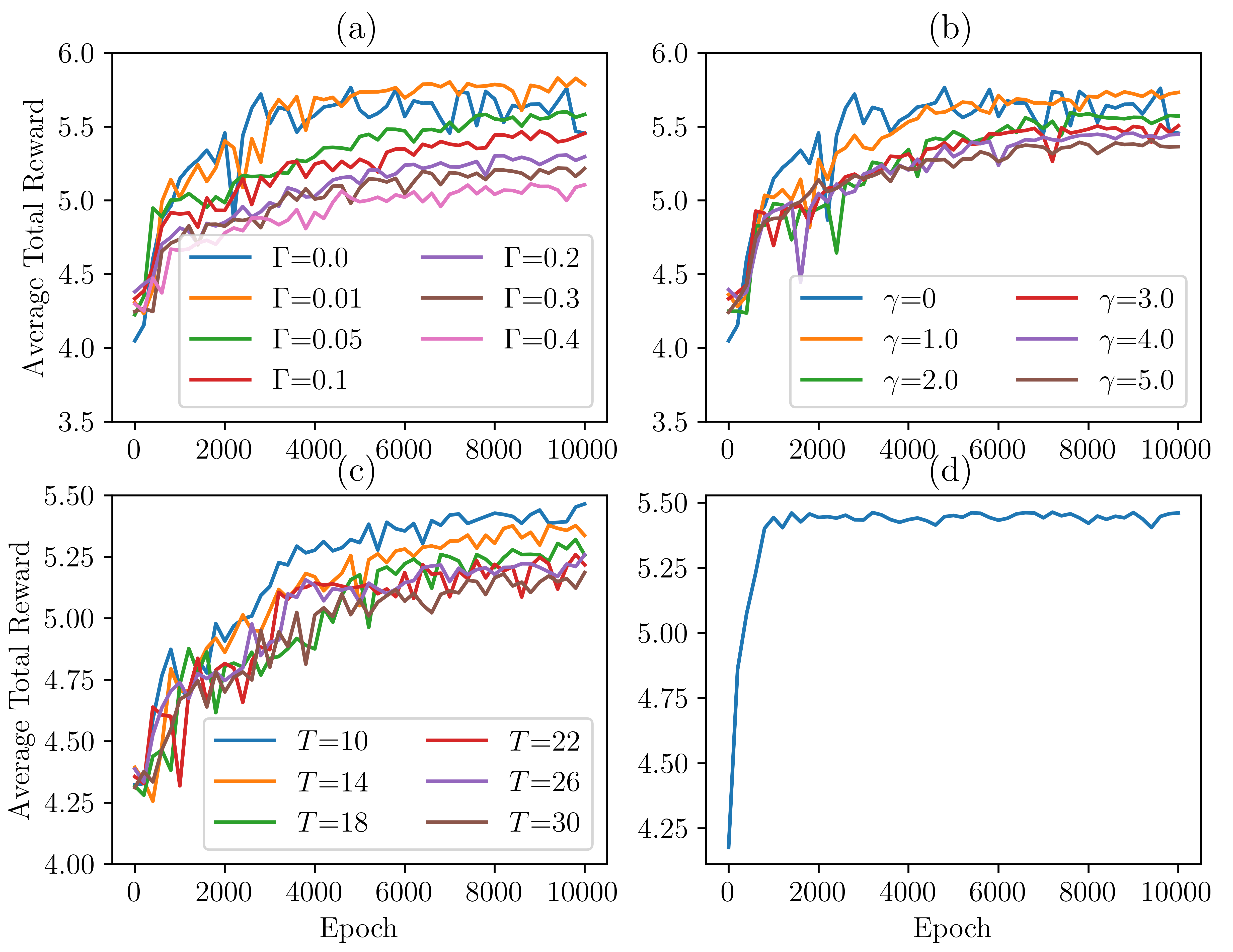}}
	\caption{The average cumulative reward on the validation set as a function of the number of training epochs during the DRL algorithm training process. (a) $\Gamma(\gamma=4, T=10)$; (b)$\gamma=4(\Gamma=0.1, T=10)$; and (c) $T(\Gamma=0.1,\gamma=4)$ for Case (2). (d) All combinations of environmental parameters for Case (3). The maximum training epoch is 10000, and the model effect is verified on the validation set every 200 epoch.}
	\label{fig:6}
\end{figure}

\begin{figure}
	\centering{\includegraphics[width=\columnwidth]{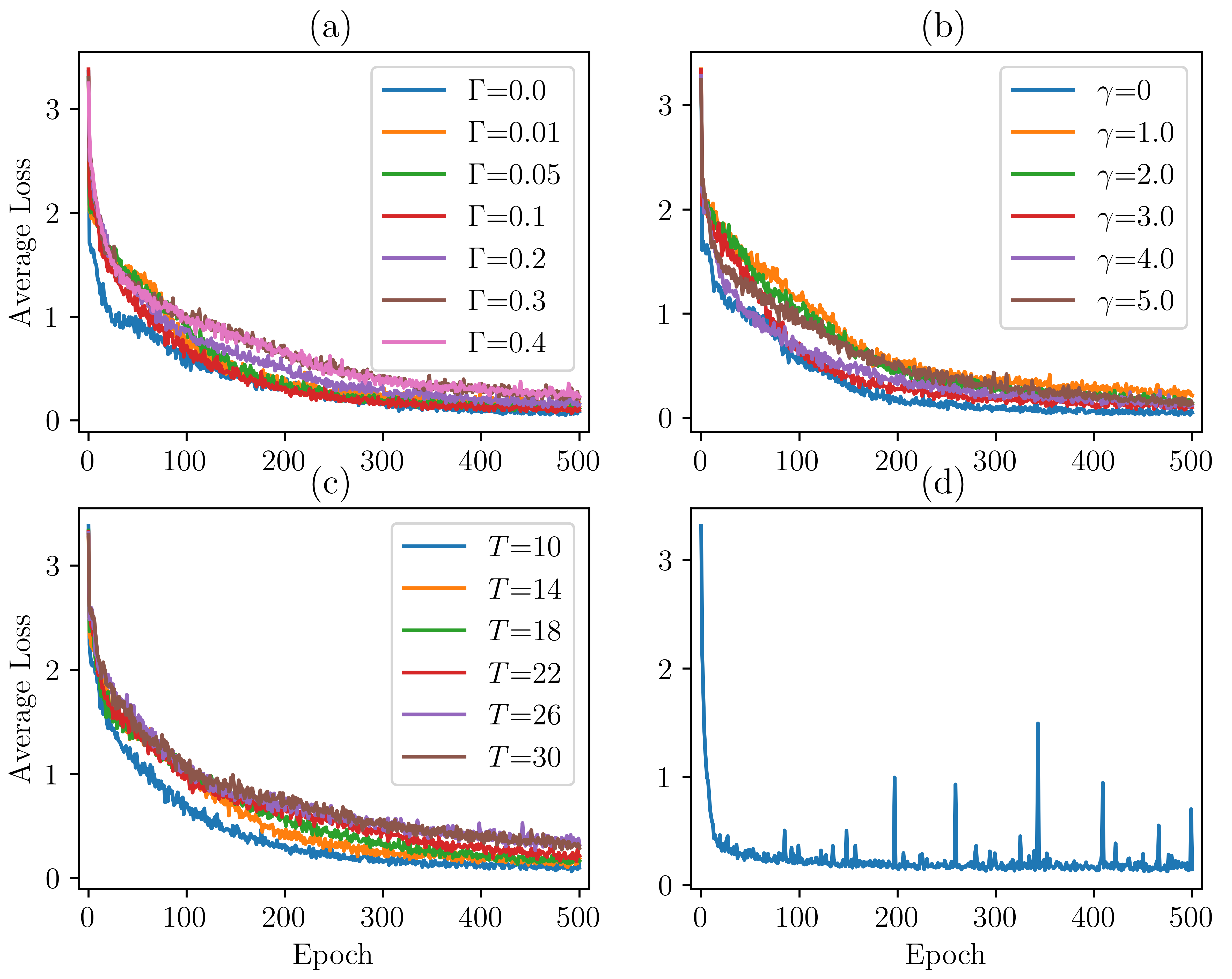}}
	\caption{The average loss on the validation set as a function of the number of training epochs during the SL algorithm training process. (a) $\Gamma(\gamma=4, T=10)$; (b)$\gamma=4(\Gamma=0.1, T=10)$; and (c) $T(\Gamma=0.1,\gamma=4)$ for Case (2). (d) All combinations of environmental parameters for Case (3).}
	\label{fig:7}
\end{figure}

Since the training logic of different algorithms varies, we use the same neural network structure and ignore differences in hyperparameters across algorithms. We focus on demonstrating the best-performing models trained by each algorithm. The hyperparameters for the two algorithms are shown in Tabs.~\ref{tab:2} and \ref{tab:3}, respectively.

For the DRL algorithm, the average cumulative reward of the validation set can be used as a criterion to determine whether the model has completed training. As shown in Fig.~\ref{fig:6}, we provide the average total reward of the DRL algorithm during the training process of Case (2) and Case (3) (checked every 200 rounds). All models trained by the DRL algorithm converge after adequate training. Fig.~\ref{fig:6} (d) illustrates the training process of Case (3). In this case, a training epoch requires training on all combinations of environmental parameters. A larger training set facilitates faster convergence of the model.

For the SL algorithm, the average loss of the validation set during the training process can be used as a criterion to determine whether the model has completed training. As shown in Fig.~\ref{fig:7}, the average loss of the SL algorithm during the training process of Case (2) and Case (3) is provided. After sufficient training, all the models trained by SL algorithm are converge. For Case (3), convergence is faster in the process of training the NN due to the use of larger data sets.

\clearpage                      

\section*{References}
\bibliographystyle{unsrt}
\bibliography{References_library.bib}

\end{document}